\pdfoutput=1

\documentclass[11pt]{article}

\usepackage{ACL2023}

\usepackage{times}
\usepackage{latexsym}

\usepackage[T1]{fontenc}

\usepackage[utf8]{inputenc}

\usepackage{microtype}

\usepackage{inconsolata}

\usepackage{graphicx}

\usepackage{amsmath}
\usepackage{amssymb}
\usepackage[skip=0pt]{caption}
\usepackage{subcaption}
\usepackage{pifont}
\usepackage{booktabs}
\usepackage{algorithm}
\usepackage{algpseudocode}
\algrenewcommand\algorithmicrequire{\textbf{Input:}}
\algrenewcommand\algorithmicensure{\textbf{Output:}}
\usepackage{algpseudocode}
\usepackage{threeparttable}
\usepackage{multirow}
\usepackage{tabularx}
\usepackage{cleveref}
\usepackage{enumitem}
\usepackage[textsize=small]{todonotes}
\usepackage{mathtools}
\usepackage{tcolorbox}
\usepackage{soul}
\usepackage{multirow}
\usepackage{colortbl}
\colorlet{lightgray}{White!30!lightgray}
\colorlet{lightblue}{White!70!MidnightBlue}

\usepackage{nicematrix}
\usepackage{arydshln}

\usepackage{listings}
\usepackage{caption}

\lstdefinestyle{papersafe}{
  captionpos=b,
  language=Python,
  basicstyle=\ttfamily\small,
  numbers=none,
  backgroundcolor=\color{code-bg},
  frame=single,
  rulecolor=\color{black},
  breaklines=True,
  breakatwhitespace=true,
  showstringspaces=false,
  keywordstyle=\bfseries\color{oi-blue},       
  commentstyle=\color{oi-green!60!black}, 
  stringstyle=\color{oi-red!85!black},         
  identifierstyle=\color{black},               
  emph={np,torch,nn,plt,fit,predict},          
  emphstyle=\color{oi-purple},
  ndkeywordstyle=\color{oi-orange},            
  aboveskip=2pt, belowskip=2pt, framesep=2pt, columns=fullflexible,
  prebreak=\mbox{\textcolor{gray}{$\hookleftarrow$}},
  postbreak=\mbox{\textcolor{gray}{$\hookrightarrow$}\space},
}
\crefname{figure}{Figure}{Figures}
\crefname{equation}{Equation}{Equations}
\crefname{appendix}{Appendix}{Appendices}
\crefname{table}{Table}{Tables}
\crefname{section}{Section}{Sections}

\newcommand{\interalia}{\emph{inter alia}}
\newcommand{\llama}{\text{Llama}}
\newcommand{\splade}{\text{SPLADE}}
\newcommand{\spladelatest}{\text{SPLADE-v3}}
\newcommand{\spladeadapt}{\text{CSPLADE}}
\newcommand{\spladeecho}{\text{CSPLADE-Echo}}
\newcommand{\spladebidirectional}{\text{CSPLADE-Bi}}

\makeatletter
\def\adl@drawiv#1#2#3{%
        \hskip.5\tabcolsep
        \xleaders#3{#2.5\@tempdimb #1{1}#2.5\@tempdimb}%
                #2\z@ plus1fil minus1fil\relax
        \hskip.5\tabcolsep}
\newcommand{\cdashlineCustom}[1]{%
  \noalign{\vskip\aboverulesep
           \global\let\@dashdrawstore\adl@draw
           \global\let\adl@draw\adl@drawiv}
  \cdashline{#1}
  \noalign{\global\let\adl@draw\@dashdrawstore
           \vskip\belowrulesep}}
\makeatother

\title{\textsc{CSPLADE}: Learned Sparse Retrieval with Causal Language Models}

\author{
Zhichao Xu,\,\,
Aosong Feng,\,\,
Yijun Tian,\,\,
Haibo Ding,\,\,
Lin Lee Cheong\\
Amazon Web Services\\
{\tt xzhichao@amazon.com}
}

\begin{document}
\maketitle
\begin{abstract}
In recent years, dense retrieval has been the focus of information retrieval (IR) research. While effective, dense retrieval produces uninterpretable dense vectors, and suffers from the drawback of large index size. Learned sparse retrieval (LSR) has emerged as promising alternative, achieving competitive retrieval performance while also being able to leverage the classical inverted index data structure for efficient retrieval. However, limited works have explored scaling LSR beyond \text{BERT} scale. In this work, we identify two challenges in training large language models (LLM) for LSR: (1) training instability during the early stage of contrastive training; (2) suboptimal performance due to pre-trained LLM's unidirectional attention. 
To address these challenges, we propose two corresponding techniques: (1) a lightweight \emph{adaptation} training phase to eliminate training instability; (2) two model variants to enable \emph{bidirectional information}. 
With these techniques, we are able to train LSR models with 8B scale LLM, and achieve competitive retrieval performance with reduced index size. 
Furthermore, we are among the first to analyze the performance-efficiency tradeoff of LLM-based LSR model through the lens of model quantization. Our findings provide insights into adapting LLMs for efficient retrieval modeling. 
\end{abstract}

\section{Introduction}
\label{sec:intro}
Recently, the main research focus in information retrieval (IR) has been on dense retrieval and related techniques~\cite[\interalia]{karpukhin-etal-2020-dense,lin2022pretrained,zhu2023large,xu2025surveymodelarchitecturesinformation}. 
Dense retrieval encodes queries and documents into high-dimensional sparse vectors. Although effective, these dense vectors are difficult for humans to interpret in terms of their semantic meanings. Moreover, encoding and storing the dense vectors for the whole document collection can be resource-intensive. For example, encoded flat index of MS MARCO passage corpus~\cite{bajaj2016ms} with \text{Llama-2-7b} dense retriever takes up 135G disk space~\cite{ma2024fine}, which is over 50 times larger than the 2.6G Lucene index from Lucene's implementation of BM25. 

To mitigate these drawbacks of dense retrieval, a different line of works investigates learned sparse retrieval (LSR). Inspired by traditional sparse retrieval models~\cite{sparck1972statistical,robertson1995okapi}, LSR encodes queries and documents into vocabulary-sized vectors with a backbone language model and the language model head, where each dimension of the vector represents the ``impact'' of the corresponding token~\cite{formal2021spladev1,formal2021spladev2,mallia2021learning}. A canonical example of LSR is \text{SPLADE}~\cite{formal2021spladev1,formal2021spladev2}. 
It encodes text with \text{BERT}~\cite{devlin-etal-2019-bert}, then applies pooling and log-saturation~\cite{fang2004formal} to ensure the resulting vocabulary-sized vector contains non-negative values in each dimension, making it suitable for use in an inverted index. 
Combined with established training methodologies in dense retrieval such as contrastive learning~\cite{oord2018representation}, hard negatives mining~\cite{karpukhin-etal-2020-dense,xiong2021ance} and knowledge distillation~\cite{hofstatter2020marginmse}, LSR has demonstrated competitive performance with \text{BERT}-style encoder-only masked language models~\cite{kong2023sparseembed,lassance2024spladev3}.

Scaling has been a winning recipe for natural language processing (NLP) and IR~\cite{kaplan2020scalinglaw,hoffmann2022chinchillascalinglaw,fang2024scalinglawsfordenseretrieval}. Recent works~\cite[\interalia]{ma2024fine,lee2024nvembedtechnicalreport,wang-etal-2024-improving-text,xu2025distillationversuscontrastivelearning,zhang2025qwen3embedding} have explored scaling dense retrieval and reranking with pre-trained decoder-only large language models (LLM) such as \text{Llama}~\cite{touvron2023llama2} and \text{Mistral}~\cite{Jiang2023Mistral7B}, which have demonstrated superior performance compared to \text{BERT} family models.
However, there has been limited effort in training LSR models beyond \text{BERT} scale, i.e., 110M and 330M.
In our preliminary experiments, we identified two key challenges in training LSR with decoder-only LLMs: 
(1) the ReLU activation function used in log-saturation of \text{SPLADE} leads to \emph{training instability} problem in early stage of contrastive training, commonly referred to as the dying ReLU problem~\cite{lu2019dyingrelu}; (2) the \emph{unidirectional attention} of decoder-only LLMs leads to suboptimal retrieval performance~\cite{lee2024nvembedtechnicalreport,behnamghader2024llm2vec}. 
To address these two challenges, in this paper we propose two techniques:
\begin{itemize}
    \item We propose a lightweight \emph{adaptation phase} training, where we adapt the pre-trained language model on unlabeled texts with a combination of causal language modeling loss and log-saturation loss. Experimental results show that as few as 10k adaptation steps can eliminate the training instability problem in subsequent contrastive training. 
    \item We explore two variants to able unidirectional LMs to capture \emph{bidirectional information}: (1) applying the echo embedding idea~\cite{springer2024echoembedding}, where we repeat the input sequence and only gather the representation from the second occurrence of text sequence; (2) directly disabling the causal language modeling (CLM) mask, and letting the language model adapt to bidirectional information in the contrastive training phase, similar to~\citet{lee2024nvembedtechnicalreport}. Our experiments show that both variants significantly improve upon causal language model with unidirectional information. 
\end{itemize}
We refer to our method as \text{Causal SPLADE} (\text{CSPLADE}). With the proposed techniques, we are able to train LSR model with up to 8B scale pre-trained LM (\text{Llama-3.1-8B}), while achieving competitive performance with only MS MARCO passage retrieval training set (41.3 MRR on MS MARCO passage retrieval, 55.3 NDCG@10 on BEIR) and reduced index size (<8G Lucene index of MS MARCO passage corpus versus 135G flat dense index). 

A significant challenge in adopting LLMs for retrieval lies in the scalability and inference latency. 
We examine several popular quantization methods such as \text{LLM.int8}~\cite{dettmers2022llmint8}, \text{torchao}~\cite{torchao}, and report the performance-efficiency tradeoff when applying on CSPLADE. 
We find while calibration-free quantization methods achieve reduced GPU memory usage, they does not necessarily lead to inference speedup in small batch sizes. 
Our findings underscore the importance of in-depth study of model quantization methods specifically designed and optimized for neural retrieval models.

\section{Background and Notations}
\label{sec:background}
In this section we introduce the task definition and notations used in this paper, and further provide background for learned sparse retrieval, with a special focus on \splade~\cite{formal2021spladev1,formal2021spladev2}.
\subsection{Task Definition and Notations}
\label{subsec:background_notations}
Given a query $Q$, the task is to find a ranked list of $k$ documents, denoted by $\{D_1, D_2, \ldots, D_k \}$, that exhibit high relevance to $Q$. Retrieval is performed by finding top-$k$ documents from document collection $\mathcal{C}$, where $|\mathcal{C}| \gg k$. We denote the retrieval model parameterized by $\theta$ as $f_{\theta}(\cdot)$.
To support efficient retrieval, the document collection is typically pre-encoded offline by the retrieval model, resulting in what is referred to as the \emph{document index}. At retrieval time, the incoming query is first encoded by the retrieval model, after which a similarity search is performed against the pre-built document index.

\subsection{Sparse Retrieval}
\label{subsec:background_sparse}
Different from the prevalent dense retrieval method~\cite[\interalia]{karpukhin-etal-2020-dense,xiong2021ance} that represents a document with a dense vector, the sparse retrieval method represents a document with a vocabulary-sized vector where most of the elements are zeros, hence the term ``sparse''. This sparse vector representation can be subsequently used in an inverted index for efficient retrieval. 
Examples of sparse retrieval include classical methods such as the boolean model~\cite{salton1984booleanlogic} and probabilistic retrieval models like BM25~\cite{robertson1995okapi}.

Traditional sparse retrieval methods focus on capturing lexical match signals, which hinders performance is finding semantically relevant documents~\cite{yates2024neurallexicalsearch}. Learned sparse retrieval emerges as a way to leverage pre-trained language models to mitigate this weakness. At a higher level, LSR can be viewed as a way to learn token importance or “impact” scores from data~\cite{dai2019deeper,bai2020sparterm,mallia2021learning}.

\subsection{\splade}
\label{subsec:background_splade}
We detail the formulation of~\splade~\cite{formal2021spladev1,formal2021spladev2}, which serves as the basis of the proposed method (\cref{sec:method}). 
Denote vocabulary as $\mathcal{V}$, a document as $D$, tokens as $\{ t_1, t_2, \ldots t_{|D|}\}$, where $t_i$ is the $i$-th token, and its corresponding contextualized representation (e.g., from pre-trained \text{BERT}) $\{\mathbf{h}_1, \mathbf{h}_2, \ldots \mathbf{h}_{|D|} \}$.
For each $\mathbf{h}_i$, we project the hidden representation to a vocabulary-sized vector $\mathbf{H}_i \in \mathbb{R}^{|\mathcal{V}|}$ with the language modeling head (e.g., masked language modeling head for \text{BERT}). The $j$-th dimension of $\mathbf{H}_i$ represents the importance of token $j$ (in vocabulary $\mathcal{V}$) to token $i$ in the input sequence, which in practice is the $\text{logit}_j$ from the LM head output.
Given $\mathbf{H}_{D} = \{\mathbf{H}_1, \mathbf{H}_2, \ldots \mathbf{H}_{|D|} \}$ of tensor shape $(|\mathcal{V}|, |D|)$,  \splade~then applies a max-pooling along the sequence length dimension, i.e., across all tokens, followed by ReLU activation and log rescaling to get the vocabulary-sized representation for the input document $d$:
\begin{equation}
    \mathbf{D} = \log \Big( 1+\text{ReLU} \big( \text{MaxPooling} (\mathbf{H}_D) \big) \Big) \in \mathbb{R}^{|\mathcal{V}|}
    \label{eq:transformation}
\end{equation}
A similar operation can also be applied to query $Q$ to get query representation $\mathbf{Q}\in \mathbb{R}^{|\mathcal{V}|}$.
Denote a similarity function as $s(\cdot)$ (e.g., dot product), we can optimize \splade~with the standard InfoNCE loss~\cite{oord2018representation} for contrastive training. Denote a training pair $(Q, D^{+})$, where $D^{+}$ is relevant to query $Q$, and $\{D_N \}$ is a list of documents not relevant to $Q$, the ranking loss is denoted by: 
\begin{multline*}
    \mathcal{L}_{rank}(Q, D^{+}, \{D_N\})
    = -\log p(D=D^+ | Q)\\
    = - \log \frac{e^{s(Q, D^+)}}{e^{s(Q, D^+)} + \sum\limits_{D_i^- \in \{D_N \}} e^{s(Q, D_i^-)}}
\end{multline*}
In practice, $\{D_N \}$ often includes hard negatives and in-batch negatives~\cite{qu-etal-2021-rocketqa,ma2024fine}.
\footnote{Subsequent works~\cite{formal2021spladev2,kong2023sparseembed} have explored other training strategies such as distillation~\cite{hofstatter2020marginmse}. In this study we opted for straightforward contrastive training, as more complex training strategies are orthogonal to the focus of this paper.}
Notice that~\cref{eq:transformation} already achieves a certain degree of sparsity by ensuring the non-negativity. In addition, \splade~also employs \texttt{FLOPs} regularization~\cite{paria2020minimizingflops} to further enhance sparsity in order to learn efficient sparse representation. 
Denote \texttt{FLOPs} regularization loss for $Q$ and $D$ as $\mathcal{L}_{reg}^Q$ and $\mathcal{L}_{reg}^D$, respectively, and $\lambda_Q$, $\lambda_D$ as the corresponding coefficients, \splade~optimizes the final loss as:
$$
\mathcal{L} = \mathcal{L}_{rank}(Q, D^{+}, \{D_N\}) + \lambda_Q \mathcal{L}_{reg}^Q + \lambda_D \mathcal{L}_{reg}^D 
$$
In practice, $\lambda_Q$ and $\lambda_D$ are tuned as hyperparameters to balance performance and efficiency.

\section{Challenges and Proposed Techniques}
\label{sec:method}

\splade's effectiveness at \text{BERT}-scale has been demonstrated by extensive prior studies~\cite[\interalia]{formal2021spladev2,formal2022spladeplus,kong2023sparseembed,li2023slim}.
However, limited studies have explored to train~\splade~beyond \text{BERT}-scale, i.e., to extend to pre-trained causal large language models like \llama~\cite{touvron2023llama2} or \text{Mistral}~\cite{Jiang2023Mistral7B} to further improve performance with stronger backbone LMs and extensive pre-training.
In our preliminary experiments where we replace \text{BERT}~\cite{devlin-etal-2019-bert} in original \splade~implementation with causal LLMs like \text{OPT-1.3B}~\cite{zhang2022opttechnicalreport} and \text{Llama-3.2-1B}~\cite{grattafiori2024llama3technicalreport}, we identified key challenges (\cref{subsec:method_challenges}). We then propose corresponding strategies to enable training \splade~with causal language models (\cref{subsec:method_proposed}). We name our method as \text{Causal SPLADE} (\text{CSPLADE}).

\subsection{Challenges}
\label{subsec:method_challenges}
First, we notice the \emph{training instability} problem in early stage of contrastive training. Recall in~\cref{eq:transformation}, the ReLU activation function is used to ensure non-negativity of the vocabulary-sized representation $\mathbf{Q}$ and $\mathbf{D}$. As the training starts, ReLU neurons quickly become inactive and only output 0 for any input. This is referred to as the dying ReLU problem in literature~\cite{lu2019dyingrelu}, and is caused because of the initialization of the parameters to be optimized, the backbone language model in our case. 
To validate this hypothesis, we also attempted to train other encoder-only models including \text{MosaicBERT}~\cite{portes2023mosaicbert} and \text{ModernBERT}~\cite{warner2024modernbert}. However, training consistently failed for the same reason, despite extensive tuning of the learning rate warmup strategy and other hyperparameters. 

Second, we observe that in the original \splade~implementation, the model first projects the contextualized representation of each token into the vocabulary space to get $|D|$ vectors. It then applies MaxPooling over the sequence length dimension to get a single vector representation. Apparently, this sequence-level MaxPooling operation becomes suboptimal for causal LLMs with \emph{unidirectional attention}, as early tokens in the input sequence cannot attend to later ones, leading to a loss of information in the final vector representation. To summarize, we believe these two challenges hinders the exploration to extend \splade~to the large-scale pre-trained casual LLMs.

\subsection{Proposed Method}
\label{subsec:method_proposed}
We have identified two challenges in~\cref{subsec:method_challenges}, namely, (1) the training instability problem and (2) the unidirectional information problem. To address these challenges, we introduce a two-phase training pipeline to get the final retrieval model: (1) \emph{adaptation phase training} to improve training stability and (2) \emph{enabling bidirectional information} in contrastive learning to capture richer context.

\paragraph{Adaptation Phase Training.}
As pointed out by~\citet{lu2019dyingrelu}, the reason for dying ReLU is due to the ill-initialized model parameters. Therefore we tackle the training instability problem by adapting the pre-trained causal language model for improved initialization to be used in subsequent contrastive training phase.
In particular, for an input sequence $D$, we can compute the output logits $\mathbf{H}_D \in \mathbb{R}^{|\mathcal{V}|\times |D|}$. We then apply the similar transformation in~\cref{eq:transformation}, except that we remove the MaxPooling operation:
$$
\mathbf{H}_D^*=\log \big( 1+ \text{ReLU}(\mathbf{H}_D) \big)
$$
Here $\mathbf{H}_D^*$ tensor has the same shape as $\mathbf{H}_D$, but all elements are non-negative. We use this non-negative tensor to compute causal language model loss to maximize the probability of the target sequence (the same input sequence in our case). 
Denote this causal language modeling loss intended for ReLU adaptation as $\mathcal{L}_{ReLU}$, and the vanilla causal language model loss as $\mathcal{L}_{\textit{CLM}}$, we optimize a final loss:
$$
\mathcal{L}_{Adapt}=\mathcal{L}_{\textit{CLM}}+ \lambda_{ReLU}\cdot \mathcal{L}_{ReLU}
$$ 
where $\lambda_{ReLU}$ is a trade-off weight balancing two loss terms, which in practice is set to 1. 
Note here the vanilla causal language modeling loss is computed directly between $\mathbf{H}_D$ and target sequence. Therefore we can compute the two loss terms with one single forward pass. 
This loss form enables us to adapt the pre-trained causal LLM for the ReLU activation used later in contrastive training. 
For ease of understanding, we show the pseudo code (PyTorch style) in~\cref{appendix:lossfunctioncode}.

This adaptation training strategy bears two strengths: (1) it can be performed on any unlabeled texts; (2) it can be efficiently combined with domain adaptation training~\cite{gururangan-etal-2020-dont} to further improve downstream task performance in the target domain.
Empirically, we find as few as 10k steps of adaptation eliminate the training instability problem, suggesting its efficacy. 

\paragraph{Enabling Bidirectional Information.}
Having addressed the training instability problem, we move on to improve learning representations. 
As we previously discussed (\cref{subsec:method_challenges}), the unidirectional nature of pre-trained causal language models hinders the capability to effectively learn sequence representations. Different works have attempted to tackle this problem in dense retrieval~\cite[\interalia]{springer2024echoembedding,lee2024nvembedtechnicalreport,behnamghader2024llm2vec}. 
Motivated by these prior efforts, we study two variants of enabling bidirectional information for pre-trained causal LLMs: (1) \emph{echo embedding}~\cite{springer2024echoembedding} and (2) \emph{directly enabling bidirectional attention}~\cite{lee2024nvembedtechnicalreport}. 

Echo embedding does not change the model architecture of pre-trained LLM, but instead changes the input. It repeats the input sequence twice, and only perform the pooling on the second occurrence of the input sequence. This way, the earlier tokens of the input sequence in the second occurrence can still attend to later tokens in the first sequence, enabling bidirectional information. On the other hand, it essentially doubles the input length, making it more computationally extensive in training and also higher latency in inference. 

We also experiment with the idea in~\citet{lee2024nvembedtechnicalreport}, where we directly enable the bidirectional attention of pre-trained causal LLM by removing the causal mask, then use the model in subsequent contrastive training. This method merges the phase of adapting for bidirectional information into contrastive training phase, and does not cause additional computational overhead compared to the echo embedding variant. 
Empirical results show that both variants significantly outperform the causal LLM baseline, with the variant using bidirectional attention yielding slightly superior performance.

We train three variants of \splade~model with \text{Llama-3}~\cite{grattafiori2024llama3technicalreport} as the backbone causal language models, although the techniques discussed can also be applied to other pre-trained causal LLMs. We refer to \splade~with unidirectional information as \text{CSPLADE} (only with adaptation training), and the two bidirectioanl variants as \spladeecho~and \spladebidirectional, respectively. 

\section{Experiment Setup}
\label{sec:experiment_setup}
In this section we discuss datasets, baselines, implementation and other experimental details. 

\subsection{Datasets}
\label{subsec:experiment_dataset}
We focus on the task of passage retrieval. We train the retrieval models on the training split of MS MARCO passage retrieval dataset~\cite{bajaj2016ms} which consists of approximately 500k training queries. We use a blend of BM25~\cite{robertson1995okapi} and \text{CoCondenser}~\cite{gao-callan-2022-cocondenser} hard negatives which is publicly available\footnote{\url{https://huggingface.co/datasets/Tevatron/msmarco-passage-aug}}.

For evaluation, we evaluate in-domain retrieval performance on the official MS MARCO passage retrieval DEV set of 6,980 queries. We also evaluate on TREC DL19 and DL20 that consist of 43 and 54 queries respectively, with in-depth annotation. We adopt the following official evaluation metrics: MRR@10 and Recall@1000 for DEV, and NDCG@10 for DL19 and DL20. 
We also include the BEIR dataset~\cite{thakur2021beir} for out-of-domain evaluation. We evaluate the official evaluation metric NDCG@10 on 13 publically available testsets in the BEIR collection. We defer more dataset details to~\cref{appendix:dataset_details}.

\subsection{Compared Methods}
\label{subsec:experiment_baseline}
We include baseline methods of both dense retrieval and sparse retrieval. 
For dense retrieval, we include \text{CoCondenser}~\cite{gao-callan-2022-cocondenser}, \text{bi-SimLM}~\cite{wang-etal-2023-simlm}, \text{SGPT}~\cite{muennighoff2022sgpt} and \text{RepLlama}~\cite{ma2024fine}. 
\text{RepLlama} is the closest dense retrieval baseline to our method as we use the same contrastive training objective and trainset. We include results from the original paper with 
\text{Llama-2-7b}~\cite{touvron2023llama2}, as well as \text{Llama-3.1-8b}~\cite{grattafiori2024llama3technicalreport} results from~\citet{zhuang-etal-2024-promptreps}\footnote{Note we skip \text{RepLlama-3-8B}'s results on BEIR as they are not reported by~\citet{zhuang-etal-2024-promptreps}.}.

For sparse retrieval, we include the classical BM25 method~\cite{robertson1995okapi}. We also include \text{SPLADE++ SelfDistil} as well as the latest \text{SPLADE-v3} variants that use complex hard negative mining and self-distillation training strategy, reported by~\citet{formal2021spladev2,lassance2024spladev3}. Lastly, we include \text{SparseEmbed}~\cite{kong2023sparseembed} as another competitive \splade~variant. For all the baselines, we use results reported in their corresponding papers.

\subsection{Implementation and Hyperparameters}
\label{subsec:experiment_implementation}
As mentioned in~\cref{subsec:method_proposed}, we examine the effectiveness of the proposed method using pre-trained \text{Llama-3} family models as the retriever backbone. We use 1B and 8B model sizes: \text{Llama-3.2-1B} and \text{Llama-3.1-8B}\footnote{\url{https://huggingface.co/meta-llama}}. 

Our implementation is based on PyTorch, Huggingface~\cite{wolf2019huggingface}, Tevatron~\cite{gao2022tevatron} and Pyserini~\cite{lin2021pyserini}'s Lucene integration. 
After the model is trained, we use Lucene to build an inverted index and do subsequent retrieval. 
For adaptation training, we adapt the pre-trained CLMs on MS MARCO passage corpus for 10K steps, with 2,048 sequence length and 32 global batch size. We employ a cosine learning rate scheduling with 1k steps warmup. We use sequence packing~\cite{raffel2020t5} technique for computational efficiency. 
For contrastive training, we use LoRA fine-tuning~\cite{hu2021lora} to balance in-domain and out-of-domain performance~\cite{biderman2024lora}. 
We use a similar training setup as \text{RepLlama}, i.e., 15 hard negatives per positive query-passage pair, together with in-batch negatives. We use 511 in-batch negatives for both 1B and 8B models, implying a global batch size of 32. We use techniques including Flash Attention 2~\cite{dao2023flashattention}, gradient checkpointing, gradient accumulation and PyTorch FSDP~\cite{zhao2023pytorchdsdp} for scalable training. 
We train 1B model for 3 epochs and 8B model for 1 epoch using cosine learning rate scheduling. 
We mainly tune four hyperparameters: learning rate, \texttt{FLOPs} regularization coefficients $\lambda_Q$ and $\lambda_D$, and LoRA rank $R$. More hyperparameter details and hardware information are deferred to~\cref{appendix:hyperparameter_details}. 
At the inference time, we merge the LoRA adapter to the backbone model, and use \texttt{bfloat16} precision for inference. 

\section{Results and Analysis}
\label{sec:result}
We discuss in-domain evaluation results (\cref{subsec:results_msmarco}) and out-of-domain results (\cref{subsec:results_beir}). We detail the setup and results of quantization evaluation (\cref{subsec:results_quantization}) and finally discuss unsuccessful attempts (\cref{subsec:results_unsuccessful}).
\subsection{In-domain Retrieval}
\label{subsec:results_msmarco}
\begin{table*}[h!]
\centering
\caption{
Performance and index size for MS MARCO in-domain evaluation. We highlight the highest performance within each section except for the index size column.
}
\label{tab:results_msmarco}
\resizebox{0.99\textwidth}{!}{
\begin{tabular}{
lrrrrrr
}
\toprule
\begin{tabular}[c]{@{}l@{}l} \textbf{Model} \\ \, \\ \end{tabular} &
\begin{tabular}[c]{@{}l@{}l} \textbf{Size} \\ \, \\ \end{tabular} & 
\begin{tabular}[c]{@{}l@{}l} \textbf{Dev} \\ \textbf{MRR@10} \\ \end{tabular} &
\begin{tabular}[c]{@{}l@{}l} \textbf{\,} \\ \textbf{Recall@1k} \\ \end{tabular} &
\begin{tabular}[c]{@{}l@{}l} \textbf{DL19} \\ \textbf{NDCG@10} \\ \end{tabular} &
\begin{tabular}[c]{@{}l@{}l} \textbf{DL20} \\ \textbf{NDCG@10} \\ \end{tabular} & 
\begin{tabular}[c]{@{}r@{}l} \textbf{Index Size} \\ \textbf{GB} ($\downarrow$) \\ \end{tabular} \\
\rowcolor{lightgray}
\multicolumn{7}{l}{\emph{Dense Retrieval}} \\
\text{CoCondenser}~\cite{gao-callan-2022-cocondenser} & 110M & 38.2 & 98.4 & 71.7 & 68.4 & 25 \\
\text{bi-SimLM}~\cite{wang-etal-2023-simlm} & 110M & 39.1 & 98.6 & 69.8 & 69.2 & 25 \\
\text{GTR-XXL}~\cite{ni-etal-2022-gtr} & 4.8B & 38.8 & 99.0 & - & - & 25 \\
\text{RepLlama-2-7B}~\cite{ma2024fine} & 7B & 41.2 & 99.4 & 74.3 & 72.1 & 135 \\ 
\text{RepLlama-3-8B}~\cite{zhuang-etal-2024-promptreps} & 8B & \textbf{42.8} & - & \textbf{74.5} & \textbf{73.9} & 135 \\ 
\rowcolor{lightgray}
\multicolumn{7}{l}{\emph{Sparse Retrieval}} \\
BM25 ($k_1=0.9, b=0.4)$  & - & 18.4 & 85.4 & 50.6 & 48.0 & 2.6 \\
\text{SPLADE++}~\cite{formal2021spladev2} & 110M & 37.8 & 98.5 & \textbf{73.6} & 72.8 & 2.6 \\
\text{SparseEmbed}~\cite{kong2023sparseembed} & 110M & 39.2 & 98.1 & - & - & - \\
\text{SPLADE-v3}~\cite{lassance2024spladev3} & 110M & \textbf{40.2} & - & 72.3 & \textbf{75.4} & 3.1 \\
\rowcolor{lightgray}
\multicolumn{7}{l}{\emph{Proposed Method}} \\
\text{CSPLADE-1B} & 1.3B & 38.2 & 98.5 & 73.2 & 68.9 & 2.6 \\
\text{CSPLADE-Echo-1B} & 1.3B & 38.8 & 98.9 & 72.9 & 69.5 & 4.6 \\
\text{CSPLADE-Bi-1B} & 1.3B & \textbf{40.4} & \textbf{99.0} & \textbf{73.8} & \textbf{69.8} & 5.6 \\
\cdashlineCustom{1-7}
\text{CSPLADE-8B} & 8B & 39.5 & 99.0 & 73.0 & 68.0 & 7.5 \\
\text{CSPLADE-Echo-8B} & 8B & 40.8 & 98.9 & 73.5 & 70.7 & 4.5 \\
\text{CSPLADE-Bi-8B} & 8B & \textbf{41.3} & \textbf{99.1} & \textbf{74.1} & \textbf{72.8} & 6.7 \\

\bottomrule
\end{tabular}
}

\end{table*}
We show the in-domain results in~\cref{tab:results_msmarco}.
For dense retrieval baselines, we find \text{RepLlama-3-8B} improves upon \text{RepLlama-2-7B}, suggesting the backbone language model's capacity is critical for retrieval performance. In terms of sparse retrieval baselines, \spladelatest~achieves the highest performance, surpassing \text{RepLlama-3-8B} on the DL20 benchmark. This observation suggests knowledge distillation from a strong cross-encoder teacher is effective for improving \text{BERT}-sized model's performance. 

For the proposed method, we find that \spladeecho~and \spladebidirectional~significantly outperforms \spladeadapt~variant. 
For example, \text{CSPLADE-Bi-1B} achieves 40.4 MRR@10 on DEV compared to \text{CSPLADE-1b}. 
This suggests the importance of enabling bidirectioanl information. 
In addition, we observe that 8B models outperform the 1B counterparts, reiterating the importance of backbone model capacity. 
Between \spladeecho~and \spladebidirectional, \spladebidirectional~performs slightly better. Although we note it also requires more careful hyperparameter tuning to achieve strong performance. 
Finally, we note that the best performing \text{CSPLADE-Bi-8B} still lags behind its dense retrieval counterpart. We hypothesize this performance gap stems from is \text{RepLlama-3-8B}'s use of a single 4,096-dimension dense vector, which offers greater representational capacity compared to \spladeadapt's sparse representation. We intend to control the size of inverted index for efficient retrieval. A more comprehensive investigation into the trade-off between retrieval effectiveness and index size is left for future work.

\begin{table*}[h!]
\centering
\caption{Results for zero-shot passage retrieval evaluation. We highlight the best performance within each section.
}
\label{tab:results_beir}
\resizebox{\linewidth}{!}{
\begin{NiceTabular}{@{}l|rr|rrrr}
\toprule
& \multicolumn{1}{c}{\text{SPLADE-v3}} & \multicolumn{1}{c|}{\text{RepLlama-2}}  & \multicolumn{1}{c}{\text{CSPLADE-Echo-1B}} & \multicolumn{1}{c}{\text{CSPLADE-Bi-1B}} & \multicolumn{1}{c}{\text{CSPLADE-Echo-8B}} & \multicolumn{1}{c}{\text{CSPLADE-Bi-8b}} \\
\multicolumn{1}{l}{\multirow{-2}{*}{Dataset}} & 110M & 7B & 1.3B & 1.3B & 8B & 8B \\
\midrule
\multicolumn{1}{l}{Arguana} & \textbf{50.9} & 48.6 & 46.7 & 45.0 & 48.1 & \textbf{48.9} \\
\multicolumn{1}{l}{Climate-FEVER} & 23.3 & \textbf{31.0} & 23.8 & 21.8 & \textbf{29.5} & 29.4 \\
\multicolumn{1}{l}{DBPedia} & \textbf{45.0} & 43.7 & 39.5 & 39.0 & \textbf{45.2} & 44.5 \\
\multicolumn{1}{l}{FEVER} & 79.6 & \textbf{83.4} & 71.3 & 74.8 & 85.2 & \textbf{86.5} \\
\multicolumn{1}{l}{FiQA} & 37.4 & \textbf{45.8} & 35.0 & 36.3 & 39.9 & \textbf{40.5} \\
\multicolumn{1}{l}{HotpotQA} & \textbf{69.2} & 68.5 & 61.0 & 62.4 & 69.4 & \textbf{69.8} \\
\multicolumn{1}{l}{NFCorpus} & 35.7 & \textbf{37.8} & 33.2 & 32.4 & \textbf{37.7} & 37.2 \\
\multicolumn{1}{l}{NQ} & 58.6 & \textbf{62.4} & 54.5 & 55.4 & 59.8 & \textbf{60.9} \\
\multicolumn{1}{l}{Quora} & 81.4 & \textbf{86.8} & 81.5 & 79.6 & 86.9 & \textbf{87.1} \\
\multicolumn{1}{l}{SCIDOCS} & 15.8 & \textbf{18.1} & 16.0 & 15.1 & 17.4 & \textbf{17.6} \\
\multicolumn{1}{l}{SciFact} & 71.0 & \textbf{75.6} & 71.1 & 71.1 & 73.2 & \textbf{73.9} \\
\multicolumn{1}{l}{TREC-COVID} & 74.8 & \textbf{84.7} & 77.7 & 71.6 & \textbf{84.0} & 83.2 \\
\multicolumn{1}{l}{Touche-2020} & 29.3 & \textbf{30.5} & 32.1 & 37.7 & 38.5 & \textbf{38.9} \\
\hline
\multicolumn{1}{l}{Average} & 51.7 & \textbf{55.1} & 49.5 & 49.4 & 55.0 & \textbf{55.3} \\ 
\bottomrule
\end{NiceTabular}
}

\end{table*}

\subsection{Zero-shot Retrieval}
\label{subsec:results_beir}

We show the zero-shot retrieval performance in~\cref{tab:results_beir}. Due to space constraints, we report results only for \text{SPLADE-v3} and \text{RepLlama-2} as representative baselines, and refer readers to~\cref{appendix:additional_results} for the full set of baseline comparisons.

We find that \text{RepLlama-2-7b} achieves better out-of-domain performance compared to other dense retrieval baselines (avg. 55.1 NDCG@10), and is also better compared to the most competitive sparse retrieval method \text{SPLADE-v3}.
Meanwhile, \text{CSPLADE-Echo-1B} and \text{CSPLADE-Bi-1B} achieve average NDCG@10 scores of 49.5 and 49.4, respectively, underperforming \text{SPLADE-v3}. This again suggests the effectiveness of distillation training especially when backbone LM's capacity is limited. 
However, when we increase the capacity of backbone language model, the performance significantly improved. For example, \text{CSPLADE-Echo-8B} and \text{CSplade-bi-8b} outperform \text{SPLADE-v3} by a large margin, and are able to achieve performance on par with \text{RepLlama-2-7b}. 
This observation suggests that the proposed method's effectiveness is also generalizable to out-of-domain zero-shot retrieval.

\subsection{Model Quantization and Latency}
\label{subsec:results_quantization}
\begin{figure*}[htp]
    \centering
    \includegraphics[width=\textwidth]{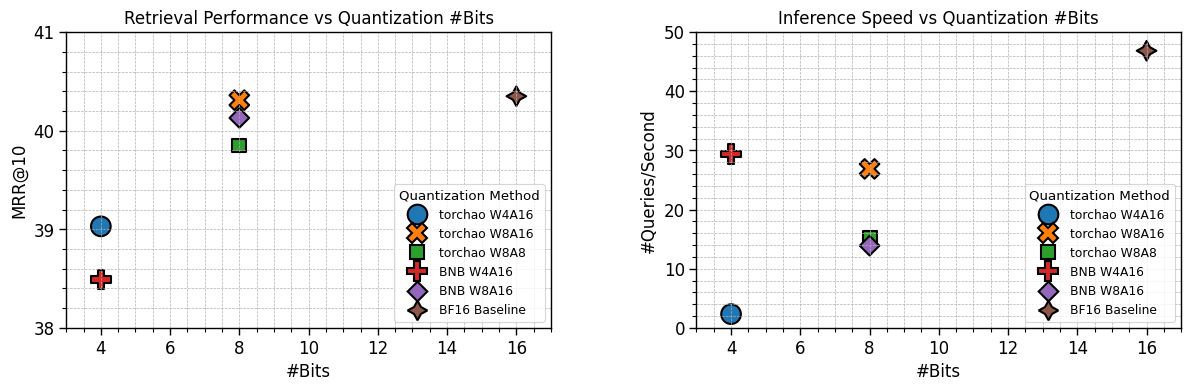}
    \caption{Quantization evaluation results for \text{CSPLADE-Bi-1B}. Left figure shows performance while right figure shows inference speed. See~\cref{appendix:quantization_details} for \text{CSPLADE-Bi-8B} results.}
    \label{fig:1b_quantization}
\end{figure*}
One big obstacle to applying LLM for real-world retrieval applications is latency. 
Therefore, we examine the efficacy of the prevalent inference optimization technique --- quantization --- in the case of learned sparse retrieval. We include two calibration-free quantization methods: \text{LLM.int8}~\cite{dettmers2022llmint8} and native PyTorch quantization implementation \text{torchao}~\cite{torchao}. For \text{LLM.int8}, we use \texttt{INT4} and \texttt{INT8} weight-only quantization in BitsAndBytes\footnote{\url{https://github.com/bitsandbytes-foundation/bitsandbytes}}. For \text{torchao}, we use \texttt{INT4} and \texttt{INT8} weight-only quantization and \texttt{INT8} weight-and-activation quantization implementation\footnote{\url{https://github.com/pytorch/ao}}. See~\cref{appendix:quantization_details} for more details about these quantization methods. 

For quantization methods and the \texttt{bfloat16} baseline, we build the inverted index using larger batch size with quantized models. Then we benchmark latency by measuring the model's speed to encode queries by setting batch size to 1. We report retrieval performance MRR@10 on MS MARCO DEV set, and use queries/second as the latency metric\footnote{We opt to not include retrieval time from Lucene index but solely focus on encoding speed from the backbone language model, as Lucene retrieval speed is the same for all quantization methods, and is mainly dependent on the index size and CPU hardware capacity.}.

We show the quantization evaluation results in~\cref{fig:1b_quantization}. First we notice that low bits quantization hurts the model's performance, e.g., 4-bit quantized models show significant performance degradation, while for 8-bit quantized models the degradation is less severe. The observation is consistent with existing works in quantization evaluation~\cite[\interalia]{dettmers2023inferencescalinglaws,xu-etal-2024-beyond-perplexity,hong2024decodingcompressedtrust}.
In terms of the inference speed, we find that at 1B scale, both quantized models actually slow down inference speed without customized GPU kernels support, compared to extensively optimized \texttt{bfloat16} baseline with Flash Attention 2 and \texttt{torch.compile}, although they do require less GPU memory. We also conduct experiments on quantization methods that require calibration, including \text{GPTQ}~\cite{frantar2023gptq} and \text{AWQ}~\cite{lin2024awq}, but find they lead to severe performance degradation, i.e., <10 MRR@10. The reason is the mismatch between the causal language modeling objective used in calibration and ranking objective used to fine-tune the retrieval model. 
To summarize, our findings highlight the need for a more careful and targeted investigation of quantization methods as well as developing efficient quantized model inference kernels tailored to retrieval models.

\subsection{Unsuccessful Attempts}
\label{subsec:results_unsuccessful}
We discuss two unsuccessful attempts throughout the experiments of adapting pre-trained causal language model for learned sparse retrieval. 

To mitigate the dying ReLU problem, we experimented with a biased reparameterization trick~\cite{wang2024nonnegative} that is inspired by Gumbel-softmax trick~\cite{jang2016gumbelsoftmax}. Here the trick is illustrated in PyTorch style:
\begin{multline*}
    \texttt{z = F.relu(z).detach() + F.gelu(z)} \\
    \texttt{- F.gelu(z).detach()}
\end{multline*}
where \texttt{z.detach()} eliminates the gradients of \texttt{z}. With this trick, the forward output still equals \texttt{F.relu(z)} but the gradient is computed with respect to \texttt{F.gelu(z)} which has gradients to negative input~\cite{hendrycks2016gaussian}. The reparameterization is biased as the gradient of GeLU is different from ReLU, which may lead to inferior performance. However, we found this trick do not fully mitigate training instability -- the training still fails in certain combinations of hyperparameters.

In contrastive training phase, we have also experimented with using full parameter fine-tuning for \text{Llama-3} instead of LoRA fine-tuning . 
We observe this leads to inferior results on BEIR datasets, while the similar problem is less significant for \text{BERT}-scale models. We hypothesize the reason is modern causal LLMs like \text{Llama}'s extensive pre-training make them prone to overfitting. We leave this overfitting problem to future investigation. 

\section{Related Works}
\label{sec:relatedwork}
In this section, we discuss existing learned sparse retrieval methods, and refer to existing survey works for dense retrieval methods~\cite{lin2022pretrained,zhu2023llm4irsurvey,xu2025surveymodelarchitecturesinformation}.
\citet{zamani2018snrm} propose \text{SNRM} to embed documents and queries in a sparse high-dimensional latent space, and enforce sparsity via $l_1$ regularization. 
\text{DeepCT}~\cite{dai2019deeper} learns to reweight terms via learning contextualized representations; but this method does not mitigate vocabulary mismatch problem as it does not employ query and document expansions. Later works further improve retrieval performance via expansion technique and corresponding aggregation mechanism, exemplified by \text{SparTerm}~\cite{bai2020sparterm}, \text{SPARTA}~\cite{zhao2020sparta} and \text{EPIC}~\cite{macavaney2020expansion}.
\splade~\cite{formal2021spladev1} note pre-trained masked language model's masked language modeling head is particularly suited for projecting contextualized representations to vocabulary space. It additionally draws inspiration from $\log(\text{tf})$~\cite{fang2004formal} and employs \texttt{FLOPs}~\cite{paria2020minimizingflops} regularization to improve performance. 
Later iterations of \splade~\cite{formal2021spladev2,formal2022spladeplus,lassance2024spladev3} further improve by switching to MaxPooling aggregation mechanism and using sophisticated training strategies including hard negatives mining and distillation from cross-encoder teachers. 
Followup works propose to enable fine-grained query-document terms interactions to better capture relevance, examplified by \text{SparseEmbed}~\cite{kong2023sparseembed}, \text{SLIM}~\cite{li2023slim} and \text{SPLATE}~\cite{formal2024splate}. 
However, these studies are still limited to pre-trained masked language models, especially \text{BERT}~\cite{devlin-etal-2019-bert} family models.
Some concurrent works have attempted to train SPLADE-style models beyond encoder backbones.
\citet{qiao2025leveragingdecoderarchitecturesforlearnedsparseretrieval} compare encoder-decoder models like \text{Flan-T5}~\cite{chung2024scalinginstructionfinetunedlanguagemodels} versus decoder-only \text{OPT}~\cite{zhang2022opttechnicalreport}. But their experiments are limited to 3B-scale models, and do not include stronger pre-trained LLMs. 
Mistral-SPLADE~\cite{doshi2024mistralspladellmsbetterlearned} train Mistral-7B~\cite{Jiang2023Mistral7B} with echo embeddings~\cite{springer2024echoembedding}.
In this work we formally examine the challenges of training learned sparse retrieval models with pre-trained LLMs as backbones, and propose corresponding mitigation strategies. 

\section{Conclusion and Future Works}
\label{sec:conclusion}
In this paper, we focus on extending learned sparse retrieval, specifically \splade, to pre-trained causal language models. We identify two challenges: the training instability problem and the unidirectional information problem. To solve these problems, we propose a lightweight adaptation training phase to eliminate training instability and design two model variants to enable bidirectional information. With these techniques, we achieve competitive performance by training \splade~with 8B scale pre-trained causal language model, while maintaining a minimized index size. Further, we analyze how model quantization affects learned sparse retrieval and discuss implications for future improvement. 

We envision two future work directions. From the training perspective, the effect of dataset scaling, in both unsupervised adaption phrase training, and subsequent supervised fine-tuning should be carefully studied~\cite{hoffmann2022chinchillascalinglaw}. Further, different training strategies, such as knowledge distillation~\cite{hinton2015distillingknowledgeneuralnetwork} and Matryoshka Representation Learning~\cite{kusupati2022matryoshka} remain to be explored.
From the inference perspective, the inference latency of casual LLMs as the retriever backbone is still prohibitive, and inference-free learned sparse retrieval~\cite{formal2021spladev1,formal2021spladev2,geng2025competitivesearchrelevanceinferencefree} is a promising future direction. How to further optimize retrieval index specifically for learned sparse retrieval is another important question~\cite{bruch2024efficient}.

\section*{Limitations}
In this work we focus on studying causal language models for learned sparse retrieval, specifically \splade~due to its high performance. We benchmarked \text{Llama-3} family models. Whether the proposed methodology applies to other backbone language model and learned sparse retrieval methods requires further investigation and benchmarking. 
The effectiveness of learned sparse retrieval on long documents should also be carefully examined.
Given the limited space, we leave further investigation of model quantization to future work. 

\section*{Ethical Considerations}
This paper focuses on modeling improvement and the experiments are conducted on public benchmarks.
To the best of our knowledge, this paper does not raise potential ethical concerns or risks. 



\bibliography{custom}
\bibliographystyle{acl_natbib}

\appendix

\label{sec:appendix}

\section{Code of Loss Function in Adaptation Training}
\label{appendix:lossfunctioncode}
We show the pseudo code in~\cref{lst:adapt_code_fullwidth}.

\begin{figure*}[th]
  \begin{lstlisting}
# Input: 
# x: input sequence of tokens, shape (batch_size, sequence_length)
# model: causal language model
# targets: target sequence, which is the same as input, shifted by 1
# loss_fn: cross-entropy loss function
# adapt_loss_factor: hyperparameter to control the balance between two terms

import torch

# Model predicts logits for each token in the sequence
logits = model(x)  # (batch_size, sequence_length, vocabulary)
shift_logits = logits[:, :-1, :]  # Shift the logits by 1
shift_targets = x[:, 1:]  # Shift the input sequence to get the target sequence

# Compute the causal language modeling loss
# Note that the tensors need to be flattened to be used in torch CrossEntropyLoss
clm_loss = loss_fn(shift_logits.view(-1, vocab_size), shift_targets.view(-1)) 

adapt_logits = torch.log(1 + torch.relu(shift_logits))
adapt_loss = loss_fn(adapt_logits.view(-1, vocab_size), shift_targets.view(-1))

loss = clm_loss + adapt_loss_factor * adapt_loss

# Return the computed loss
return loss
  \end{lstlisting}
  \caption{Pseudo code for adaption phase training loss computation.}
  \label{lst:adapt_code_fullwidth}
\end{figure*}

\section{Dataset Details}
\label{appendix:dataset_details}
Four of the datasets we used in experiments (NFCorpus~\cite{boteva2016full}, FiQA-2018~\cite{maia2018fiqa}, Quora\footnote{\url{https://www.kaggle.com/c/quora-question-pairs}}, Climate-Fever~\cite{diggelmann2020climate}) do not report the dataset license in the paper or a repository.
For the rest of the datasets, we list their licenses below:
\begin{itemize}[leftmargin=*]
    \item MS MARCO~\cite{bajaj2016ms}: MIT License for non-commercial research purposes.
    \item ArguAna~\cite{wachsmuth2018retrieval}: CC BY 4.0.
    \item DBPedia~\cite{hasibi2017dbpedia}: CC BY-SA 3.0.
    \item FEVER~\cite{thorne2018fever}: CC BY-SA 3.0.
    \item HotpotQA~\cite{yang2018hotpotqa}: CC BY-SA 4.0.
    \item NQ~\cite{kwiatkowski-etal-2019-natural}: CC BY-SA 3.0.
    \item SCIDOCS~\cite{cohan2020specter}: GNU General Public License v3.0.
    \item SciFact~\cite{wadden2020fact}: CC BY-NC 2.0.
    \item TREC-COVID~\cite{voorhees2021trec}: "Dataset License Agreement".
    \item Touche-2020~\cite{bondarenko2020overview}: CC BY 4.0.
\end{itemize}

\section{Hyperparameter Details}
\label{appendix:hyperparameter_details}
\begin{table*}[t]
\vspace{0pt}
\centering
\caption{Hyperparameters used in training \spladeadapt. We use 5\% of the total training steps for learning rate warmup. Global BZ denotes global batch size..
}
\label{tab:hyperparameters_passage}
\begin{tabular}{
lrrrrrrr
}
\toprule
Model & LR & LR Scheduler & \#Epochs & Global BZ & LoRA Rank & $\lambda_{Q}$ & $\lambda_{D}$ \\
\midrule
\text{CSPLADE-1B} & 5e-5 & Cosine & 3& 32 & 64 & 0.003 & 0.003 \\
\text{CSPLADE-Echo-1B} & 5e-5 & Cosine & 3 & 32 & 64 & 0.003 & 0.003 \\
\text{CSPLADE-Bi-1B} & 5e-5 & Cosine & 3 & 32 & 64 & 0.003 & 0.003 \\ 
\midrule
\text{CSPLADE-8B} & 1e-4 & Cosine & 1 & 32 & 16 & 0.03 & 0.03 \\
\text{CSPLADE-Echo-8B} & 1e-4 & Cosine & 1 & 32 & 16 & 0.03 & 0.03 \\
\text{CSPLADE-Bi-8B} & 1e-4 & Cosine & 1 & 32 & 16 & 0.03 & 0.03 \\

\bottomrule
\end{tabular}
\end{table*}
We show details of hyperparameters in~\cref{tab:hyperparameters_passage}. We mainly tune four hyperparameters: learning rate, LoRA Rank, \texttt{FLOPs} regularizer coefficient $\lambda_Q$ and $\lambda_D$. We notice that increasing LoRA Rank did not improve retrieval performance, but led to performance degradation on BEIR datasets, therefore we use rank=16 for 8B models. 1B models are trained on a single EC2 p4d.24xlarge intance with 8xA100 40GB GPUs, while 8B models are trained on two p4d.24xlarge instances. 

\section{Additional Experiment Results}
\label{appendix:additional_results}

We show additional BEIR results in~\cref{tab:results_beir_appendix}.

\begin{table*}[h!]
\centering
\caption{Additional results for zero-shot passage retrieval evaluation. We highlight the best performance within each section.
}
\label{tab:results_beir_appendix}

\begin{tabular}{lrrrrrr}
\toprule
& \multicolumn{1}{c}{\text{BM25}} & \multicolumn{1}{c}{\text{SPLADE++}}  & \multicolumn{1}{c}{\text{SparseEmbed}} & \multicolumn{1}{c}{\text{SGPT}} & \multicolumn{1}{c}{\text{CSPLADE-1B}} & \multicolumn{1}{c}{\text{CSPLADE-8B}} \\
\multicolumn{1}{l}{Dataset} & - & 110M & 110M & 5.8B & 1.3B & 8B \\
\midrule
\multicolumn{1}{l}{Arguana} & 39.7 & \textbf{52.5} & 51.2 & 51.4 & 44.7 & \textbf{45.2} \\
\multicolumn{1}{l}{Climate-FEVER} & 16.5 & 23.0 & 21.8 & \textbf{30.5} & 19.5 & \textbf{27.2} \\
\multicolumn{1}{l}{DBPedia} & 31.8 & 43.6 & \textbf{45.7} & 39.9 & 39.2 & \textbf{41.8} \\
\multicolumn{1}{l}{FEVER} & 65.1 & 79.3 & \textbf{79.6} & 78.3 & 73.3 & \textbf{82.3} \\
\multicolumn{1}{l}{FiQA} & 23.6 & 34.8 & 33.5 & \textbf{37.2} & 33.2 & \textbf{39.5} \\
\multicolumn{1}{l}{HotpotQA} & 63.3 & 68.7 & \textbf{69.7} & 59.3 & 63.6 & \textbf{66.3} \\
\multicolumn{1}{l}{NFCorpus} & 32.2 & 34.8 & 34.1 & \textbf{36.2} & \textbf{36.5} & 35.7 \\
\multicolumn{1}{l}{NQ} & 30.6 & 53.7 & \textbf{54.4} & 52.4 & 52.9 & \textbf{58.8} \\
\multicolumn{1}{l}{Quora} & 78.9 & 83.4 & \textbf{84.9} & 84.6 & 81.0 & \textbf{87.7} \\
\multicolumn{1}{l}{SCIDOCS} & 14.9 & 15.9 & 16.0 & \textbf{19.7} & 15.8 & \textbf{17.0} \\
\multicolumn{1}{l}{SciFact} & 67.9 & 70.2 & 70.6 & \textbf{74.7} & 71.2 & \textbf{72.2} \\
\multicolumn{1}{l}{TREC-COVID} & 59.5 & 72.7 & 72.4 & \textbf{87.3} & 68.4 & \textbf{79.2} \\
\multicolumn{1}{l}{Touche-2020} & \textbf{44.2} & 24.5 & 27.3 & 25.4 & 34.8 & \textbf{38.0} \\
\midrule
Average & 43.7 & 50.5 & 50.9 & \textbf{52.1} & 48.8 & \textbf{53.1} \\ 
\bottomrule
\end{tabular}

\end{table*}

\section{Quantization Evaluation Details}
\label{appendix:quantization_details}

\subsection{Quantization Methods}
We experimented with quantization methods that require calibration, including \textsc{AWQ}~\cite{lin2024awq} and \textsc{GPTQ}~\cite{frantar2023gptq}, but find they led to significant performance degradation due to the misaligned objectives in model fine-tuning and calibration. We opted to focus on calibration-free quantization methods.

\subsection{Hardwares}
The inference speed is measured on single A100 GPU with 40GB memory. We use \texttt{torch.compile} and Flash Attention 2 for the \texttt{bfloat16} baseline.

\subsection{Results for 8B Models}
\begin{figure*}[htp]
    \centering
    \includegraphics[width=\textwidth]{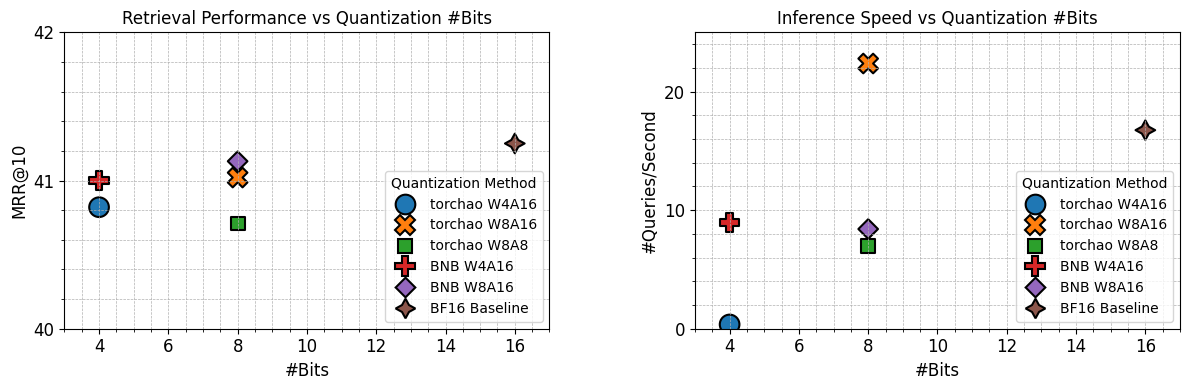}
    \caption{Quantization evaluation results for \text{CSPLADE-Bi-8B}. Left figure shows performance while right figure shows inference speed. }
    \label{fig:8b_quantization}
\end{figure*}
We report the quantization results for 8B models at~\cref{fig:8b_quantization}. We notice that at 8B scale, performance degradation is less severe compared to 1B models, and torchao W8A16 quantization improves inference speed compared to \texttt{bfloat16} baseline.

\end{document}